\documentclass[twocolumn,amsmath,showpacs,amssymb,aps,nofootinbib,floatfix]{revtex4-1}
\usepackage{epsfig,amsmath,amssymb}
\usepackage{dcolumn,color}
\usepackage{bm}
\usepackage{graphicx}

\newcommand{\observable}{N_{part}^{-1} d^2 N/d\eta^2}
\newcommand{\lns}{\ln \left( \frac{\sqrt{s}}{m_p} \right)}

\newcommand{\sqrts}{\sqrt{s}}

\newcommand{\beq}{\begin{equation}}
\newcommand{\eeq}{\end{equation}}
\newcommand{\eqcomma}{\phantom{A},\phantom{A}}
\newcommand{\order}[1]{ \mathcal{O} \left( #1 \right) }
\newcommand{\ave}[1]{\left\langle #1 \right\rangle}
\newcommand{\abs}[1]{\left| #1 \right|}

\begin{document}
\title{Limiting fragmentation as an initial-state probe in heavy ion collisions}
\author{Kayman J. Gon\c{c}alves$^1$, Andre V. Giannini$^{2,3}$, David D. Chinellato$^1$, Giorgio Torrieri$^1$}
\affiliation{$^1$ IFGW, Unicamp, Campinas, Brasil\\ $^2$ Akita International University, Yuwa, Akita-city, Japan\\ $^3$ IFUSP, USP, Sao Paulo, Brazil}
\begin{abstract}
We discuss limiting fragmentation within a few currently popular phenomenological models.
We show that popular Glauber-inspired models of particle production in heavy ion collisions, such as the two-component model, generally fail to reproduce limiting fragmentation when all energies and system sizes experimentally available are considered.
This is due to the energy-dependence of number of participants and number of collisions.   We quantify this violation in terms of the model parameters.
We also make the same calculation within a Color Glass Condensate scenario and show that the dependence of the saturation scale on the number of participants generally leads to violation of limiting fragmentation.
We further argue that wounded parton models, provided the nucleon size and parton density vary predominantly with Bjorken $x$, could in principle reproduce both multiplicity dependence with energy and limiting fragmentation.
We suggest, therefore, that an experimental measurement of deviation from limiting fragmentation in heavy ion collisions, for different system sizes and including the experimentally available range of energies, is a powerful test of initial state models.
\end{abstract}
\maketitle
\section{Introduction}
The phenomenon of limiting fragmentation in hadronic collisions was 
originally both experimentally seen \cite{busza1,busza2} and
theoretically explained at the origins of the QCD theory of strong
interactions \cite{yen}.

\begin{figure*}[t]
      \epsfig{width=0.99\textwidth,figure=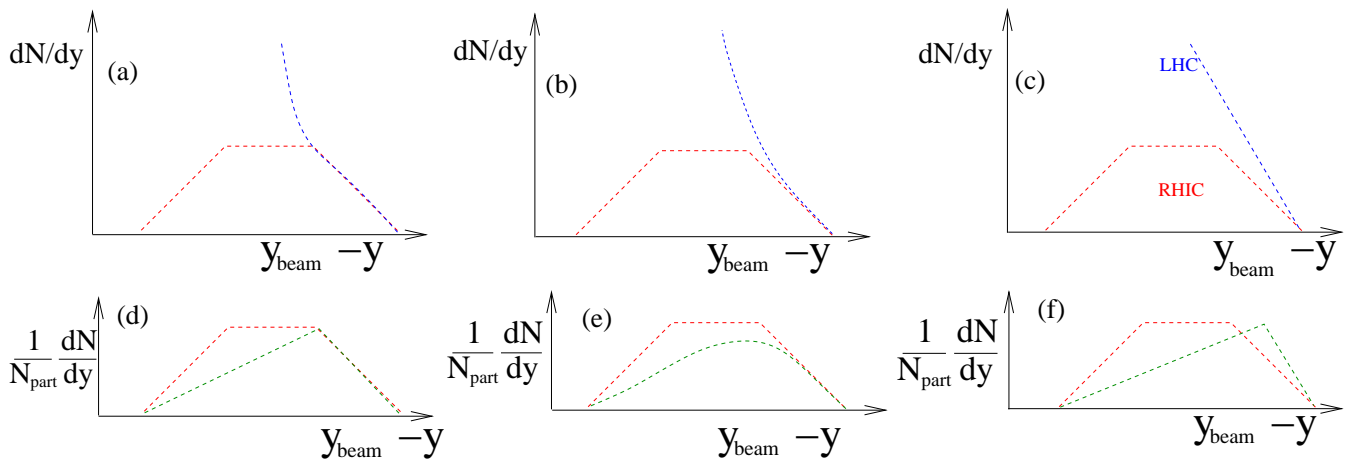}
\caption{\label{limfraglhc} (color online)
Possible limiting fragmentation scenarios when LHC and lower energies are compared (top row) and an extended comparison with different system sizes (panels d,e,f).   Case (a,d) presupposed an extended limiting fragmentation scenario up to the mid-rapidity plateau.   Case (b,e) is a smooth breaking of scaling interpolating between a universal fragmentation regime and the central rapidity plateau, case (c,f) is wholesale violation, both at central rapidity and in the fragmentation region.
The bottom row panels d,e,f show the three cases when a symmetric AA collision is compared to an asymmetric pA or dA collision.}
\end{figure*}

The definition of limiting fragmentation is that
\begin{equation}
\label{limdef}
  \left. \frac{d^2 N}{d \kappa^2}\right|_{\kappa=y-y_0} = C \eqcomma \frac{\sqrts}{C} \frac{dC}{d\sqrts}\ll 1 \eqcomma y_0 =  \ln \frac{\sqrts}{m_p}
\end{equation}
Where $C \simeq 0.65$ is an energy-independent constant and $m_p$ the proton mass.  A qualitative illustration of limiting fragmentation is shown on the  panel (a) of Fig. \ref{limfraglhc}.   As it happens well-away from mid-rapidity, both distributions in rapidity $y$ and pseudo-rapidity $\eta$ can be used for limiting fragmentation (In fact the experimental data discussed later was collected in pseudo-rapidity.  See the appendix on a discussion of this issue).

When plotted against $y-y_0$ (the difference between rapidity and beam rapidity), multiplicity distributions away from mid-rapidity should fall on a universal curve, independent of center of mass energy.   Using the second derivative of the rapidity distribution, as defined in Eq. \ref{limdef}, provides a dimensionless quantitative estimator which should vanish within error bar if limiting fragmentation holds, be $\gg 1$ if it breaks down, and is relatively insensitive to rapidity/pseudorapidity differences (see appendix).
Using directly $C$ of Eq. \ref{limdef} allows for a nice graphic comparison, while the logarithmic derivative $\frac{dC}{d\ln\sqrts} $ of  makes large fluctuations blow up, but allows a quantitative comparison across different energy regimes.

The basic explanation for limiting fragmentation is to see it as a consequence of the parton model and Bjorken scaling \cite{halzen}.   The distribution functions of parton g, i.e. the parton amplitudes in the infinite momentum frame of the nucleon, depend on the nucleon momentum in a very particular way
\begin{equation}
\left| <N(p)|g(q)> \right|^2_{p \gg m_N} \sim f(x,\ln Q^2) \eqcomma x=q/p
  \end{equation}
where $x$ is the momentum fraction and $Q^2$ the renormalization group momentum scale, corresponding to the momentum transfer of the process measuring $f(x,Q^2)$.
As the parton mass is negligible its  momentum rapidity $y_g$ is related to $x$ very simply, as
\begin{equation}
\label{ygdef}
y_g = \pm \ln \left(1/x \right)
  \end{equation}
Because of asymptotic freedom, the dependence on $Q^2$ as well as any hadronization effects  do not change the final momentum much. As a result, away from mid-rapidity fragmentation does not change the rapidity of the hadron with respect to the parton.
Thus, if one assumes parton interactions are local in rapidity, and the resulting hadron is not shifted in rapidity with respect to the parton, in other words (the delta function is simply the momentum conservation in the longitudinal axis)
\begin{equation}
\label{multgen}
  \frac{dN}{dy} \sim \int dx_A dx_B f(x_A) f(x_B) \mathcal{N}(x_A\sqrt{s}/2,x_B \sqrt{s}/2) \times
\end{equation}
\[\    \delta\left( \sinh \ln(1/x_A) + \sinh \ln(1/x_B) - \sinh y \right)
  \]
  limiting fragmentation follows naturally, since asymptotic freedom and the longitudinal momentum conservation ensure that when $x_A \gg x_B$, corresponding to $y$ very different from zero, a universal curve for $dN/dy$ depending only on $x_A \simeq e^y$ emerges.
  
  This reasoning is appropriate if the only dimensionful scale relevant to hadronic scattering is the nucleon size, i.e. if only one type of collisions, in a not-too varied range of energies is considered.
  Since then it has however been seen in a wide range of hadronic \cite{phobos,phobosprl,phobosrev,whitebrahms} and even nuclear \cite{jeon} collisions.
Indeed, a variety of approaches, ranging from Landau hydrodynamics \cite{steinberg} to the Color Glass Condensate \cite{busza,jamal,stasto,andre1,andre2} to phenomenological transport models such as  ``AMPT'' \cite{ampt0} (whose limiting fragmentation is investigated in  \cite{ampt}),
string percolation \cite{capella,palh1,palh2} and relativistic diffusion \cite{wols} are at least roughly compatible with it.

  However, limiting fragmentation turned out to work for a wider spectrum of energies, rapidities and system sizes than initially suspected.  At Relativistic Heavy Ion Collider (RHIC) energies (5-200 GeV), it seems to hold \cite{busza,phobos,jeon} for all system sizes and all energies up to an energy-dependent rapidity interval of $\order{1}$ or so.
This can be surprising.   Close to mid-rapidity, at the highest energy density, a lot of additional effects, from high energy fragmenting jets to soft gluon effects to viscous entropy production, are expected to contribute particles in a way that evolves non-trivially with energy.  The reason why energy-specific dynamics should converge to a universal distribution away from mid-rapidity is not immediately clear, since Bjorken scaling means remnants of these processes should be present away from mid-rapidity.

At RHIC energies, there was no problem to incorporate this extended limiting fragmentation into phenomenological models \cite{capella}, if not theories.   A qualitative explanation links
 \cite{weakly,surprises} limiting fragmentation to the close-to-logarithmic energy-dependence of multiplicity $dN/dy$ per participant $N_{part}$ at mid-rapidity.
The logarithmic dependence of the rapidity width (the base of the distribution) is fixed by kinematics.
Then,  ``extra parton sources'' at mid rapidity \cite{weakly,capella} produce strings connected to rapidity edges, ensuring the self-similarity of the total multiplicity distribution.

 The Large Hadron Collider (LHC) seems to have added some new twists to this story when energies of several TeV were reached.  It has conclusively deviated from logarithmic scaling of multiplicity w.r.t. energy (see for example \cite{alice}), at the same time convincingly breaking the pure number of participants scaling \cite{glabrev,glauber} of particle production.
 This has motivated the development of more complicated characterizations of the Glauber initial state, generally based on ``multi-component scenarios''.
 For instance, core-corona models posit that the event is characterized both by a ``soft medium'' (participants that underwent multiple collisions) as well as a few hard collisions (a ``corona'' of singe-hit participants).
 More phenomenologically \cite{alice,fernando,corecorona} one could assume wounded nucleons and collisions provide different admixtures to the multiplicity.

In view of these ``new developments'', it is worth revisiting limiting fragmentation at this energy.   It should first be noted that it has {\em not} as yet been confirmed or falsified experimentally \cite{alicey,cmsy,atlasy}, partially because LHC experiments have not as yet explored rapidity regimes high enough to be compared even with RHIC energies.
In principle, the three possible scenarios are described in the panels of Fig. \ref{limfraglhc}.  One can continue to have perfect limiting fragmentation in the whole region away from mid-rapidity (panel(a) ), limiting fragmentation can smoothly be achieved as rapidity increases (panel (b)) or it can just totally break down (panel (c)).  The bottom panels d,e,f show the corresponding possible scenarios in pA collisions once the distribution is normalized by the number of participants (note that only one side can limiting fragment since pA is rapidity asymmetric).

In this work, we  argue that such an investigation is indeed necessary.
We  show that the most-commonly used model for the initial state of heavy ion collisions, the Glauber model, will generally break limiting fragmentation in the parameter space where it fits LHC data {\em provided all energies and system sizes are considered}.   We show that the same is true for Color Glass Condensate (CGC) models, as implemented in \cite{andre1,andre2}.

We then argue that a currently popular model capable of bringing the applicability of limiting fragmentation to LHC energies and varying system sizes, {\em provided it is indeed confirmed experimentally}, is a ``wounded parton model'', with Glauber wounded partons \cite{woundq1,woundq2,woundq3,woundq4,woundq5} smeared in rapidity space according to Bjorken scaling.
We conclude with some experimental considerations regarding this  observable.
\section{Limiting fragmentation and the Glauber model}
The two component model is a physically intuitive parametrization of multiplicity \cite{glauber,glabrev}.  The basic idea is that, in a nucleus-nucleus collision some particles are produced in hard scatterings as a result of fragmentation of parton-parton interactions.  The rest of the particles, typically the majority, arise from ``wounded nucleons'', nuclei that underwent a collision, emitted color charge, and then emitted particles due to non-perturbative color neutralization\footnote{ We note in passing that event by event fluctuations and higher cumulants of these two mechanisms are expected to be very different.   $N_{coll}$ produced events will reflect the fluctuations inherent in the fragmentation function.   Color neutralization naturally accomodates  the negative binomial distribution, the distribution of
''throws'' you get when you throw until you get an outcome.  If parton
generation is ''locally'' not-color neutral, therefore, one needs a
negative binomial number of generations before you get a color-neutral
state.   Thus, an  energy and system size investigation of fluctuation scaling, in particular the so-called Koba-Nielsen-Olsen scaling, would parallel the limiting fragmentation investigation outlined here.   This is left to a forthcoming work.
}.   An alternative but related picture is the ``core-corona'' model, where wounded nucleons form ``the medium'' (presumably a quark-gluon plasma), while individual nucleon-nucleon collisions (and nuclei which underwent just one collision) can be thought of as being similar to proton-proton collisions \cite{corecorona}.

Thus, quantitatively, the two sources of particles are related by the parameter $f$, controlling the input of the number of collisions $N_{coll}$ vs. number of participants $N_{part}$ to the total number of ``ancestor particle sources'' $N_{anc}$:
\begin{equation}
  \label{fpar}
N_{anc}=f N_{part} + (1-f) N_{coll}\,.
\end{equation}
Physically, $N_{coll}$ scaling reflects the fragmentation of partons produced in a QCD collision.  $N_{part}$ reflects the products ``wounded nucleons'' which emit partons as they color-neutralize.

To describe multiplicity we need another
 parameter $\alpha$, controlling how many particles are produced per ancestor, interpolating between logarithmic ($\alpha=0$) and power-law ($\alpha>0$) limits
	\begin{equation}
	\label{alphadef}
	\left. \frac{dN}{dy} \right|_{y=0} = \mathcal{N}_0 N_{anc} \int_1^{\sqrts} \zeta^{\alpha-1} d\zeta 
	\end{equation}
where $\mathcal{N}_0$ is an overall normalization parameter, uncorrelated with $f,\alpha$.

In \cite{fernando} and preceding literature, it was assumed that $\alpha$ is the same for $pp$, $pA$ and $AA$ collisions.   This means one knows $\alpha$ from $pp$ fits and can obtain $f$ from $AA$ fits.
In other words,
The different exponents seen experimentally in these systems are due entirely from the fluctuation of $N_{part}$ and $N_{coll}$ of a two-component model.

This, while possible, is not guaranteed.  It is well known (The ``EMC effect'') from $eA$ scattering that parton distribution functions significantly change in a nuclear medium \cite{emc}.  Given that different energy regimes probe different $x$, it is plausible that the difference between proton and nuclear medium could give rise of different excitation functions of $dN/dy$ on top of geometric effects.  This, however, means that the $f$ parameter and the $\alpha$ parameter need to be understood together within a fit to experimental data, since the Hessian matrix relating them will contain a diagonal term.

A phenomenology tool comes from the fact that the cross-sectional area depends on energy, with its dependence  roughly
\begin{equation}
\sigma = A_1 \lns + A_2 \left( \lns \right)^2
\end{equation}
with $A_1=25$ and $A_2=0.146$ fitted from data such as \cite{sigma} \footnote{Note that this form was chosen because it provides a good interpolation of the experimental data ($\chi^2=0,97$) and should not be taken as physically motivated.}

It then becomes clear that in a general Glauber model, the ratio of $N_{coll}/N_{part}$ will depend on energy as well.
\begin{figure}[h]
\begin{center}
      \epsfig{width=0.48\textwidth,figure=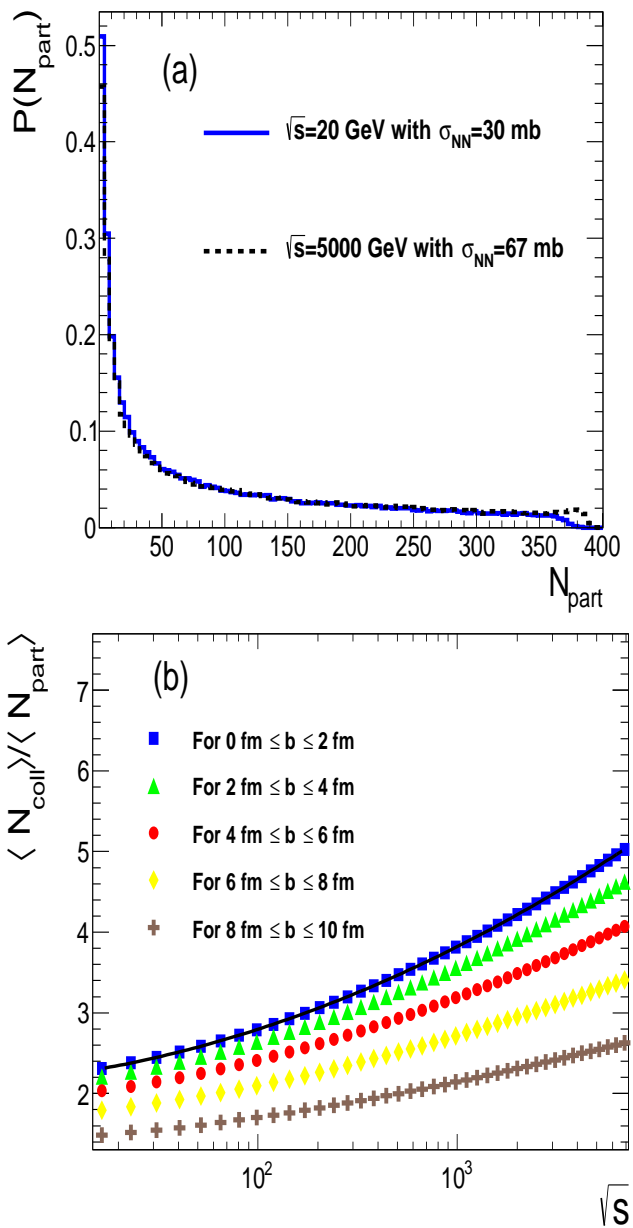}
\caption{\label{npartcoll} (color online)
Panel (a): The distribution of $N_{part}$ as a function of energy. Panel (b): The subsequent evolution of $N_{coll}/N_{part}$ in a Glauber model, with the best fit to the most central collisions added as a line}
\end{center}
\end{figure}
Indeed, a calculation using the model described in \cite{phobos,fernando} is shown in Fig. \ref{npartcoll}.  While the number of participants varies very little with energy (as is obvious, since only variations in higher order moments are permitted by definition), the number of collisions varies systematically with the increase of the inelastic scattering cross-section.

The best fit, also shown in the figure, can be parameterized by
\begin{equation}
\label{ncp}
N_{cp}(\sqrts)=T \left( \lns \right)^2+G\lns+J,
\end{equation}
Where $T,G,J$ are fit parameters (See footnote 1.  The $\chi^2$ here is negligible, $\sim 10^{-5}$).  For the top centrality bin of Au-Au, used later in multiplicity fits, their numerical values are
$T=0.043$, $G=-0.045$ e $J=2.1$

While, as panel (a) shows, the event by event $N_{part}$ distribution is relatively constant in energy (unsurprisingly, since it is bounded between zero and 2A globally, and to a good approximation to the local transverse density at each point in the transverse area), the ratio of $N_{coll}/N_{part}$ has a monotonic increase.
It is intuitively clear that in a general two-component model this will introduce an energy dependence on $dN/dy$ which does not appear in rapidity, since both $N_{part}$ and $N_{coll}$ are parameters characterizing the whole event, rapidity independent.

\begin{figure*}[t]
      \epsfig{width=0.99\textwidth,figure=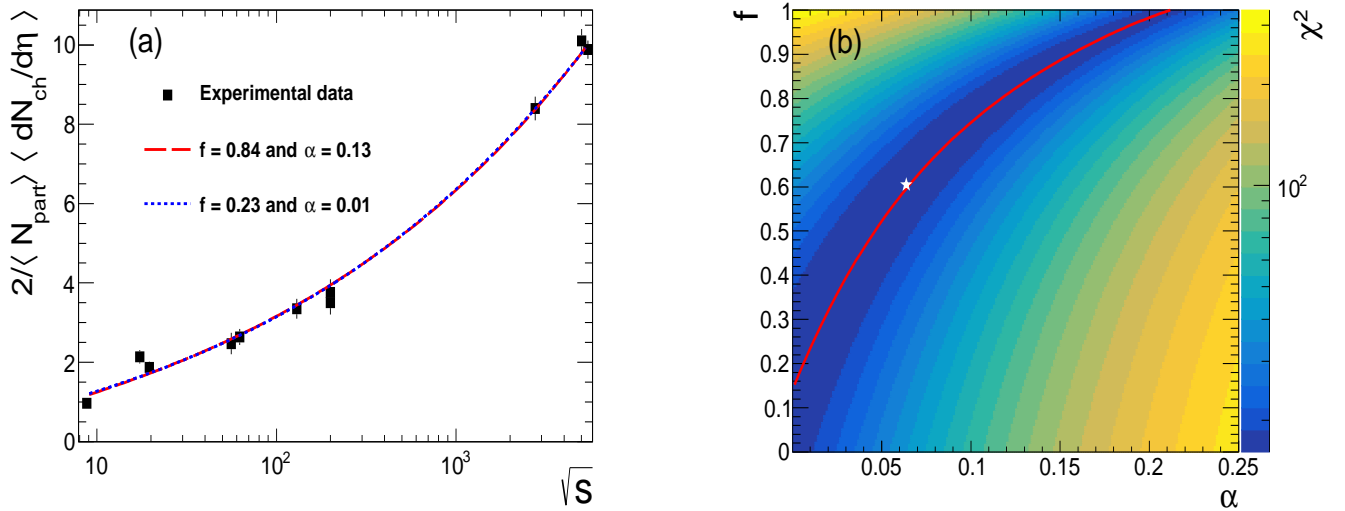}
\caption{\label{falpha} (color online)
Panel (a): experimental data, referenced in \cite{alice} with the opposite ends of the fitted curve. Panel (b): Contour of fit to midrapidity experimental data in $\alpha-f$, with best fit line highlighted and a star denoting the $\chi^2$ minimum. 
}
\end{figure*}

It is clear that $\alpha$ and $f$ are strongly correlated by experimental data.
We therefore made a $\chi^2$ contour using the data of \cite{alice} to fit them.  The results are shown in Fig. \ref{falpha}, panel (b).   We highlight the curve going through the major axis of the ellipse in $f-\alpha$ space, parametrizeable by
\begin{equation}
\label{ffit}
f_{fit}(\alpha)=F_1\alpha^4+F_2\alpha^3+F_3\alpha^2+F_4\alpha+F_5, 
\end{equation}
where the fitted values are $F_{1...5}=-250.1,175.8,-53.6,9.9,0.14$ (See footnote 1.  $\chi^2 \sim 10^{-2}$).

The $\mathcal{N}_0$ parameter of course is also correlated with $f,\alpha$. In terms of $\alpha$ its minimum can be parametrized as
\begin{equation}
  \label{defn}
  \mathcal{N}_0 \simeq N_1 \alpha^2 + N_2 \alpha + N_3
  \end{equation}
where $N_{1..3}=-3.35,1.32,0.27$ (See footnote 1.  The $\chi^2\sim 10^{-3}$).

We also note that the inclusion of both in a simultaneous fit greatly widens the space of $f-\alpha$ allowed, and in fact the fit seems to prefer a value of $f$ considerably lower ($N_{coll}$ dominated ancestors) than most of the work of this kind.

Panel (a), showing the fit quality of the opposite ends compared to experimental data, however, demonstrates one must not take this too seriously (although it can also give a decent fit of eccentricities, see for instance \cite{urom}) since
the correlation between the fit parameters (as well as the overall normalization) is such that the two extremes do a very similar job of fitting the data and in fact generate a nearly identical curve.
This, and the disagreements at lower energy also make the $\chi^2$ contour deviate significantly from the Gaussian.  

Given these ambiguities, our strategy, rather than focusing on inferring a tighter bound on $f-\alpha$ space, is to use this fit to define a relation between the two variables, based on $f_{fit}(\alpha)$, that fixes the mid-rapidity curve, and investigate limiting fragmentation given this relation.
We therefore use the graph where Eq. \ref{ffit}  was used to parametrize $f_{fit}(\alpha)$ we now scan through $\alpha$ to check how limiting fragmentation behaves for the class of Glauber models fitting mid-rapidity.

\begin{figure*}[t]
      \epsfig{width=0.99\textwidth,figure=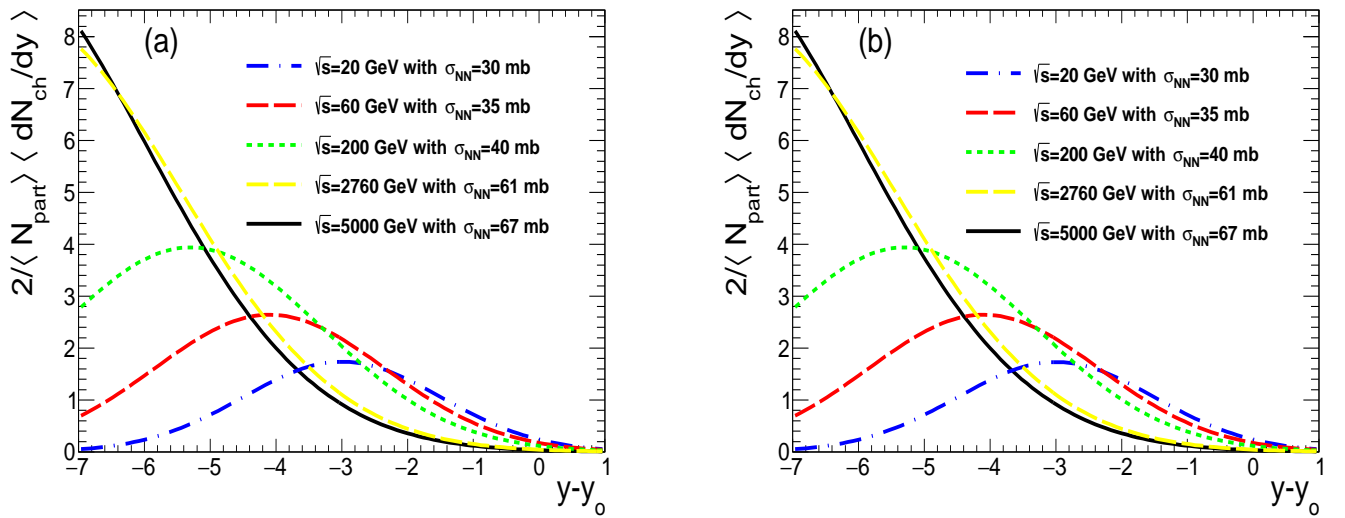}
\caption{\label{gaussians} (color online)
  Gaussian rapidity dependencies with the width given by a Landau-like kinematic parametrization, and the height extrapolated from pairs of parameters $f,\alpha$ adjusted to reproduce data at mid-rapidity
  }
\end{figure*}

First, we use the opposite ends of $(f,\alpha)$ parameter space to investigate the behavior away from mid-rapidity of $dN/dy$, assuming Gaussian distributions in a manner similar to \cite{ampt}, with widths given kinematically and heights by the observed mid-rapidity.   As can be seen in Fig \ref{gaussians}, at both extremes limiting fragmentation is broken by a similar manner.

To quantify this violation further, we used a Gaussian, a trapezoidal and a triangular distribution in $\eta$, with the bottom base given by kinematics $\Delta \eta_{bottom} = \ln (\sqrts/m_p)$ , while the top base as universally a fraction either $\Delta \eta_{top}=0$ (for a triangle and a Gaussian) or \textcolor{black}{$\Delta \eta_{top}=\Delta \eta_{bottom}/2$} of the bottom base (for a trapezium).
The height is chosen by the two-component model calculation.

\begin{figure}[hbt!]
  \begin{center}
    \includegraphics[width=7.8cm]{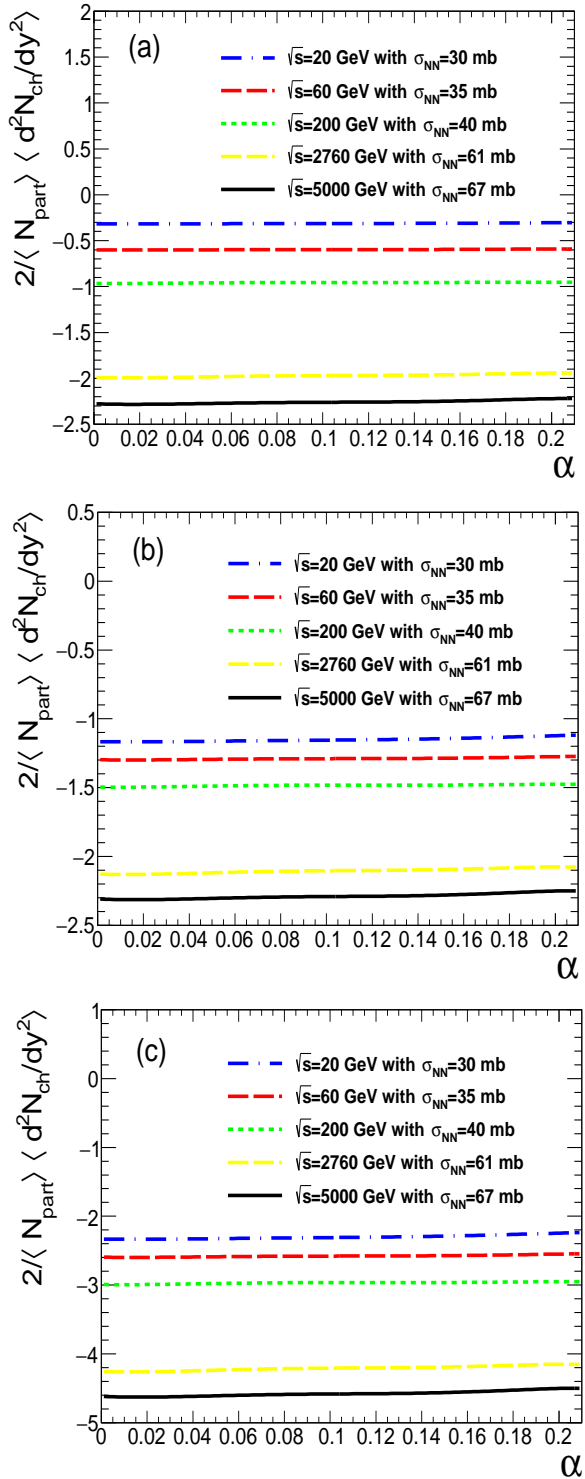}
    \end{center}
  \caption{\label{d2ndy2} (color online) The gradient of the rapidity distribution at $y-y_0=3$ assuming a Gaussian (Panel (a)), triangular (Panel (b)) and trapezoidal (Panel (c)) shapes respectively, against the $\alpha$ parameter, with the corresponding $f=f_{fit}(\alpha)$ parameter adjusted to reproduce data at mid-rapidity (Fig. \ref{falpha}). }
\end{figure}

The results are shown in Fig. \ref{d2ndy2}.   As can be seen, all cases exhibit practically constant and sizable violations of limiting fragmentation for all values of $(f,\alpha)$.  The constancy is not a surprise, given the good fit of the central rapidity for all values of $\alpha,f_{fit}(\alpha)$
Furthermore, the quantitative amount for the violation of limiting fragmentation is to a good extent independent of both the value of $\alpha$ and the model used.  We can therefore conclude with some confidence this violation is approximately what can be expected in any distribution respecting mid-rapidity multiplicity and kinematic constraints.
 Note that, while very simple, these distributions have the same functional form as the pseudorapidity distributions measured in \cite{alicey}.
The triangular distribution also has the advantage to generate limiting fragmentation analytically at $\alpha \rightarrow 0,f\rightarrow 1$, while the Gaussian one reproduces it to a good approximation for $\alpha=1/4,f=1$ \cite{steinberg}.

This limit which is not captured in our plot, since it is very far from the best fit line: $f \rightarrow 1$ prefers higher numbers of $\alpha$, which breaks limiting fragmentation.  $\alpha\rightarrow 0$ prefers very low $f$ (collision-dominated multiplicity) which again breaks limiting fragmentation.
We can therefore make the qualitative statement that a general Glauber model cannot realistically fit mid-rapidity multiplicity without a significant break in limiting fragmentation.  This is experimentally testable.

It is very simple to understand the universality of the violation of the limiting fragmentation, and generalize from Gaussians and Trapeziums (where one can get a quantitative analytically calculable answer) w.r.t. any rapidity shape, where the difference should be within a factor $\order{1}$ of the one calculated here.

All we need to assume, both are crucial assumptions for the Glauber model of particle production as described here, is
\begin{description}
\item[(i)] An emission function $F(y)$ which is specific to each ancestor.  Different emission functions $F_{part}(y)$ and $F_{coll}(y)$ would not change the result
  \begin{equation}
    \label{emission}
    \frac{dN}{dy} = \sum_{N_{anc}} F(y) \sim N_{part} F_{part}(y)+ N_{coll} F_{coll}(y)
    \end{equation}
  \item[(ii)] The base of the emission function is kinematically determined to be $\sim \lns$
  \end{description}
This section makes it clear that, as long as $N_{coll}/N_{part}$ evolves non-logarithmically and $f \ne 0$ (as is mandated by experimental data), $\observable$ cannot be universal.  These statements will hold as long as assumptions (i) and (ii) are correct, and putting in reasonable distributions shows violations big enough to be qualitative rather than small corrections.
In particular, the middle scenario in Fig. \ref{limfraglhc} (smooth approach to limiting fragmentation) is excluded by assumption (i) and by the inability of $f=1$ models to fit midrapidity data, since for this to happen the rapidity distributions of individual ancestors must be aware at how the number of ancestors (producing the mid-rapidity peak) increases.

\begin{figure}[h]
      \epsfig{width=0.5\textwidth,figure=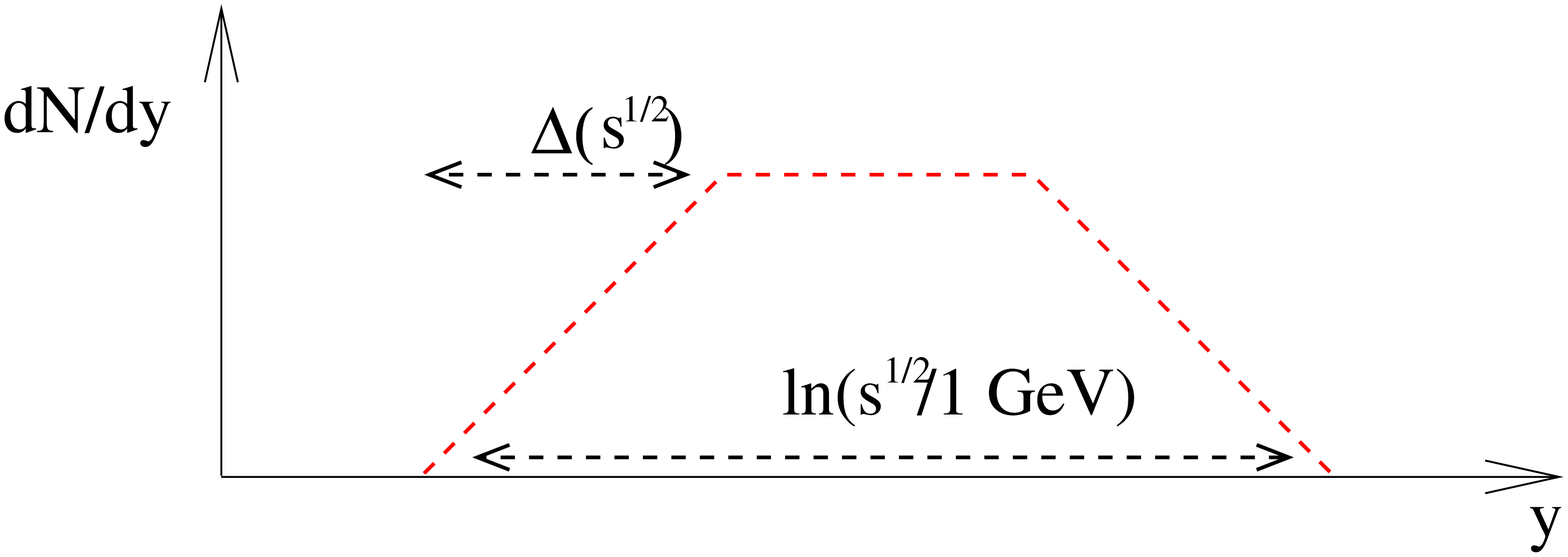}
\caption{\label{trapezium} A trapezoidal rapidity distribution with a varying width parameter
}
\end{figure}

Let us illustrate this by extending our analytically solvable model: We take the trapezium distribution (Fig. \ref{d2ndy2} panel (b)) but impose an arbitrary variation with $\sqrt{s}$ of the top rapidity plateau $\Delta(\sqrts)$ (Since this variation is arbitrary, this includes the case of $N_{part}$ and $N_{coll}$ having different plateaus (Fig. \ref{trapezium}).
Calling $N_{cp}(\sqrts)$ is the ratio of participants to collisions (shown in Fig. \ref{npartcoll} and Eq \ref{ncp} ), we will have
	\begin{equation}
	  \left. \frac{d^{2}N}{dy^{2}} \right|_{\left| y-y_0\right| \ll y_0} = \frac{Z_1}{Z_2}
          \end{equation}
        \[\   Z_1=        -4\mathcal{N}_0 N_{part} \left( f+(1-f) N_{cp}\left(\sqrts\right) \right) \left(\sqrt{s}^{\alpha}-1\right) \]
          \[\ Z_2 = \alpha\left(1-\Delta\left(\sqrt{s} \right) \right)\ln \left( \frac{\sqrts}{m_p}\right) \]
Since the only things that are permitted to vary with energy are $N_{cp}$ and $\Delta$, this means that for limiting fragmentation to occur we must have
\begin{equation}
  \label{equivalency}
  F_1 \left( f,\alpha,N_{cp}\left(\sqrts\right)  \right) = C F_2\left( \Delta\left(\sqrts\right) \right)
\end{equation}
where $C$ is the constant defined in equation \ref{limdef} and
	\begin{equation}
	F_{1}\left( f,\alpha,\sqrts \right)=\frac{\mathcal{N}_0 N_{part}}{\alpha} \times
	\end{equation}
\[\ \times \left(f+\left(1-f\right)N_{cp}\left(\sqrt{s}\right)\right) (\sqrt{s}^{\alpha}-1)  \]
        \textcolor{black}{
	\begin{equation}
		F_2\left( \Delta\left(\sqrts\right) \right) =-\frac{1}{4} 
	\left(1-\Delta\left(\sqrt{s} \right) \right)  \ln\left( \frac{\sqrts}{m_p}\right)
	\end{equation}
	}
 This is physically impossible.  Not only because the first function depends on nuclear geometry and the cross-section area and the second just on the partonic structure, but because the first depends on $\alpha$ and the second does not.

 For instance, let us suppose that the $\Delta$ evolution arises from particle distributions produced in $N_{part}$ and $N_{coll}$ having different widths  (i.e. the fragmentation region is different for wounded nuclei than for parton-parton collisions, a wholly reasonable scenario), parametrized as $\Delta_{1,2} \lns$ where $\Delta_{1,2}$ are fixed coefficients.

This would mean
	\begin{equation}
	\Delta(\sqrts) = \frac{\mathcal{N}_0 N_{part} \left(\sqrts^\alpha-1\right)}{\alpha} \times
	\end{equation}
        \[\   \times  \left( \frac{f}{\Delta_1} + \frac{(1-f)}{\Delta_2} N_{cp}(\sqrts) \right) \]
this will lead to the supposed ``constant`` $C$ of the form
\textcolor{black}{
	\begin{equation}
	C=\frac{-4\mathcal{N}_0 N_{part}\left(\sqrt{s}^{\alpha}-1\right)}{\alpha \ln \left( \frac{\sqrts}{m_p} \right)}\times
	\end{equation}
	\[\
	\times	\frac{\left(f+\left(1-f\right)N_{cp}\left(\sqrt{s}\right)\right)}{\left(1-\left(\frac{\mathcal{N}_0 N_{part}\left(\sqrt{s}^{\alpha}-1\right)}{\alpha}\right)W\left(\Delta_1,\Delta_2,f,\sqrts\right)\right)}
	\]
where
\[\ W\left(\Delta_1,\Delta_2,f,\sqrts,\right)= \frac{f}{\Delta_{1}}+\frac{\left(1-f\right)}{\Delta_{2}}N_{cp}\left(\sqrt{s}\right) \]}

\begin{figure}[hbt!]
\begin{center}
    \includegraphics[width=7.8cm]{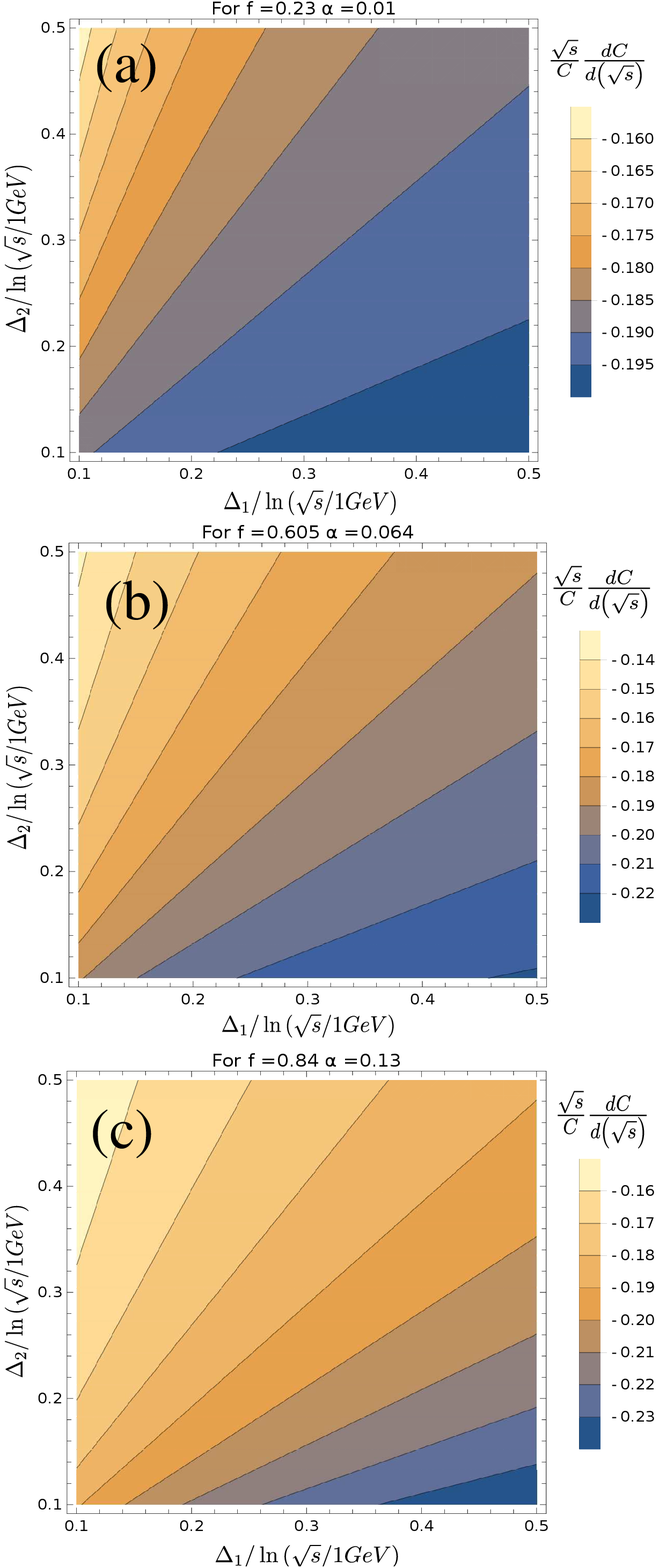}
\caption{\label{fdelta} (color online)
A scan of the expected limiting fragmentation violation in conformal units, $\sqrts C^{-1} dC/d\sqrts$ as a function of the best 
fit parameters in rapidity data $\alpha,f_{fit}(\alpha)$ (kept fixed) as well as $\Delta_{1,2}$ (on the axes) the widths of
$N_{part}$ and $N_{coll}$.  The derivative was taken at $\sqrts=$200 GeV. The three panels show the choices of $f,\alpha$
corresponding to the extremes of Fig. \ref{falpha} (panel (a) and (c)) and the $\chi^2$ minimum (panel (b)).  We verified that intermediate values are well described by a smooth interpolation between these two }
\end{center}
\end{figure}

A plot of  $\sqrts C^{-1} dC/d\sqrts$, for $f=f_{fit}(\alpha)$ and $\alpha$ adjusted according to Fig. \ref{falpha}, $\mathcal{N}_0$ given by the correlated values Eq. \ref{defn} and $\Delta_{1,2}$ scaled by $\lns$ is shown in Fig. \ref{fdelta}.    It confirms the universality of breaking of limiting fragmentation within the Glauber model, by showing that any reasonable value of $\Delta_{1,2}$ does not satisfy Eq. \ref{equivalency} for an $f,\alpha$ given by the best fig parametrization Eq. \ref{ffit}.
It is easy to see that similar equivalency, albeit with additional rapidity dependence, will arise for a Gaussian and any other shape. 
A comparison between $pA$ and $AA$ for the same number of participants (as in Fig. \ref{limfraglhc})  would be even more problematic since for $pA$ $N_{coll} \simeq N_{part}-1$ while for $AA$ no such constraint is present.

\begin{figure*}[t]
\begin{center}
      \epsfig{width=1.0\textwidth,figure=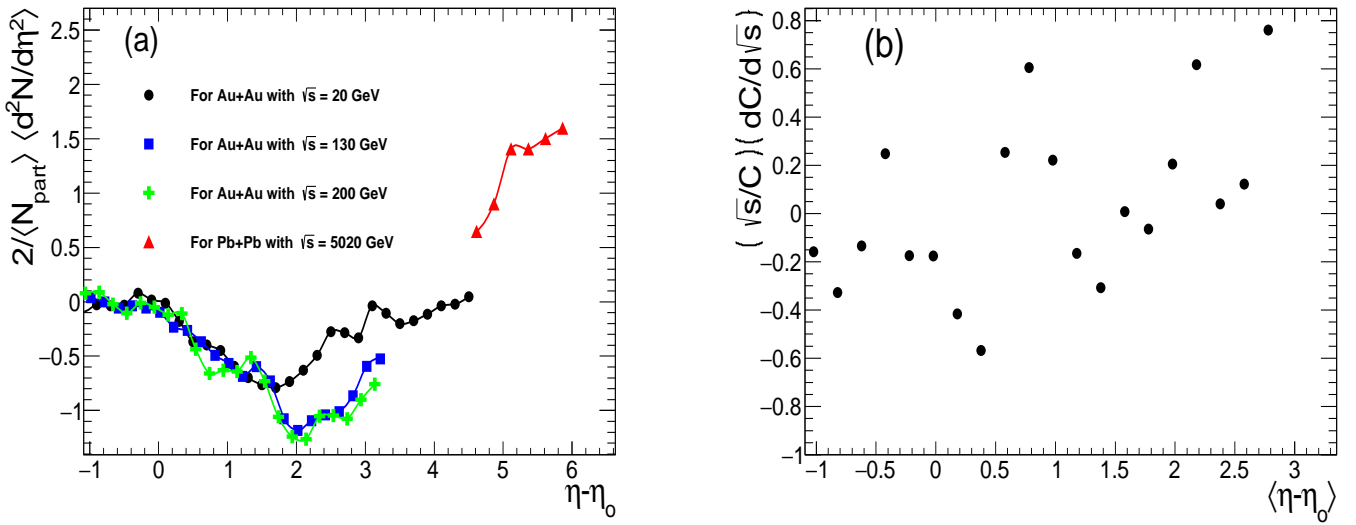}
\caption{\label{limfragexp} (color online) Panel (a): $d^2N/d\eta^2$ and the observable in Fig. \ref{fdelta} obtained from experimental data, from the compilation in \cite{jeon} (data from \cite{phobosprl}) and \cite{alicey}. Panel (b) shows the numerical derivative calculated between 130 and 200 GeV.
}
\end{center}
\end{figure*}

We also show (Fig. \ref{limfragexp}) a compedium of estimates of $d^2N/d^2\eta$ and the observable seen in Fig. \ref{fdelta}, obtained with numerical derivatives of experimental data compiled in \cite{jeon} from \cite{phobosprl} and for energies of 130 and 200 GeV (where $dN/dy$ per participant is still approximately logarithmic with $\sqrt{s}$).  Panel (a) of the graph appears to show, by eye, that limiting fragmentation is broken only weakly, if at all.   Panel (b), where a common normalization with theory was included, shows a breaking comparable to Fig. \ref{fdelta} in magnitude, but, unlike Fig. \ref{fdelta}, spread around zero.

This figure is included as a proof-of-concept, but extreme care should be taken to interpret it as a quantitative limit.  The data it is based on does not
include error bars  (\cite{jeon} and references ), and the multiplication by $\sqrt{s}$, necessary to define a dimensionless observable quantifying breaking,
magnifies fluctuations invisible to the naked eye in Fig. \ref{limfragexp}'s panel (a).  The fact that the points oscillate chaotically around zero strongly
suggest that Fig. \ref{limfragexp}'s panel (b) is in fact dominated by statistical fluctuations, and quantitatively the violation of limiting fragmentation until the top RHIC energies is parametrically smaller than that of Fig. \ref{fdelta}.    A reanalysis such as that of Fig. \ref{limfragexp} by the members of the experimental collaborations, with access to raw data, and together with LHC data
could potentially verify this claim.

Additionally, as was already noted in \cite{jeon}, extrapolating to LHC energies (Fig. 8,10 of that work) means that any quantitative disagreement with limiting
fragmentation seen at RHIC would be amplified by the departure from the logarithmic energy dependence per participant at LHC energies (predicted by a number of models in \cite{jeon} and confirmed).   A logarithmic derivative, such as the one we defined ($\sqrts d/d\sqrts =d/d\ln \sqrts$), allows to highlight the quantitative violation of limiting fragmentation inherent in this effect despite the large energy gap between top RHIC and bottom LHC energies.

Given these considerations, Fig. \ref{fdelta} and Fig. \ref{limfragexp} strongly motivate a precise quantitative estimate of the violation of limiting fragmentation both across LHC and RHIC energy regimes.

The persistence of limiting fragmentation, especially in the LHC energy regime, would preclude the system to be a superposition of nucleon-nucleon collisions.  However, such a picture was called into doubt in the literature by models where partonic degrees of freedom appear at the nuclear level, such as the Color Glass and the wounded quark model.   In the next section we describe what such models imply for limiting fragmentation.
\section{Partonic nuclear dynamics and limiting fragmentation}
It is obvious from the discussion in the previous section that any $N_{coll}$ dependence on multiplicity is inherently incompatible with limiting fragmentation.
Ultimately, the last section can be summarized in a simple statement:  It is an experimental fact that the cross-sectional area changes with energy, and this by itself will introduce an energy-dependent contribution which, unless unnecessary fine-tuning is introduced, will never seep into rapidity dependence.

All Glauber model analyses so far, however, show that some $N_{coll}$ dependence is unavoidable if nucleon-nucleon collisions are extrapolated from $pp$ to $AA$.
One answer to this is the ``wounded quark model'' \cite{woundq1,woundq2,woundq3,woundq4,woundq5} (more accurately described as wounded {\em parton} model), which also could be justified from a dynamics such as that of \cite{bgk}.

It is unsurprising that wounded parton models have more leeway in fitting multiplicities with only participant scaling, since they by definition have a greater number of parameters: Very little is known about configuration space parton distribution of the nucleons (the ``three constituent quarks'' picture is an obvious simplification at high energies).   Thus, one can substitute the effect of varying $f,\alpha$ in the two component model with parameters describing the parton transverse and longitudinal distribution within the nucleon.
The works cited above, in their own way, essentially accomplished this.

\begin{figure*}[t]
      \epsfig{width=0.85\textwidth,figure=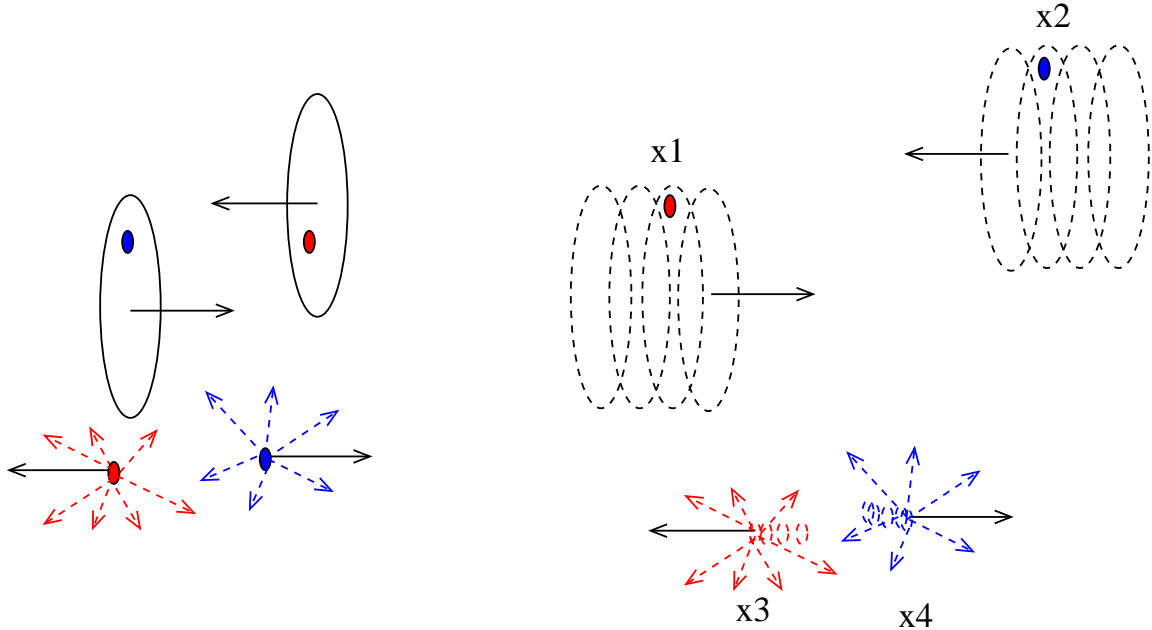}
\caption{\label{models} An illustration of the Glauber model based on wounded nucleons vs wounded partons.  The latter has the distinction that each initial degree of freedom is at a different Bjorken $x$, and shifts to a different $x$ (i.e. final rapidity bin) once ''wounded'' }
\end{figure*}

One can however also note that the fundamental difference here is that the ``size of the nucleon'', i.e. the Fourier transform of the nucleon form factor, is something that is allowed to vary with Bjorken $x$.   Boost-invariance, built into the parton model, means that space-time rapidity of the parton's trajectory and momentum rapidity are equal
\begin{equation}
\label{spacetime}
 y = -\ln \left( x^{-1} \right) = \frac{1}{2} \ln \left( \frac{z+t}{z-t} \right)  
\end{equation}
As fig \ref{models} shows, one can imagine a nucleon nucleon collision within such a model as a superposition of a ``train'' of collisions, of transverse shapes located in different bins in Bjorken $x$.

Each participant starts from that Bjorken $x$, is shifted by one or more collisions to a different Bjorken $x$, and finally emits hadrons according to some distribution in rapidity determined by this final Bjorken $x$ and fragmentation dynamics.
The transverse nucleon shape at each $x$ can also vary, and hence can 
mimic the variation of the ``nucleon-nucleon cross-section'' with energy via the growth of the typical $x$ with $\sqrts$.  Crucially, {\em the same variation} will occur in rapidity.

We  investigate this idea with a toy model, combining a Glauber-type wounded parton model with a transverse quark size parameterized by $Q_s(x)$. Note that ``s'' in $Q_s$ here might stand for ``size'', not for saturation, i.e. it might represent a nucleon-specific rather than transverse space-specific scale.
All that we need is that $Q_s$ is allowed to vary with Bjorken $x$.     Basically, we assume ``a tube of pancakes'', each at a momentum and spacetime rapidity $e^{\pm x}$ of transverse size $Q_s(x)^{-1}$.

Let us briefly recap as to how this setup generates limiting fragmentation.
The key assumptions relating limiting fragmentation to the parton model are:
\begin{description}
\item[(a)] All transverse momentum scale (denoted here by $k$) dependence is via $(k/Q_s(x))^2$, a scale (reflecting saturation or size) depending only on $x$, usually $\sim x^{-\lambda}$.  
\item[(b)] Normalization of $\tilde{T}(k,x)$ depends only on $x$ (possibly via $Q_s(x)$) or be constant
    \[\  \int^{Q^2} \frac{d^2 p_T}{p_T^2} \tilde{T}(p_T,x) \sim \mathcal{N}(x)  \]
  \item[(c)] Asymptotic freedom.  Most partons are produced with little shifts in momenta.   As a result, if $-\ln x$ is not $\gg 1$, one can assume $x \sim e^{-y}$ of the produced parton.  In simple language, asymptotic freedom prevents partons away from mid-rapidity to fragment away from their rapidity bin
\end{description}
Let us illustrate how these assumptions {\em generically} give limiting fragmentation in pp collisions.
Extending the Glauber model formalism to such a system is simple
\begin{equation}
  \label{Tax}
  T_{A}(\mathbf{s})=\int \rho _{A}(\mathbf{s},z_{A})dz_{A}  \rightarrow T_A(\mathbf{s},x) \sim F(x,Q_s(x))
\end{equation}
With $Q_s(x)$ characterizing the nucleon size at that $x$.

Now the usual definition for $T_{AB}$ can be duly updated to
\begin{equation}
 T_{AB}(\mathbf{\hat{b}},y) = \int T_{A}(\mathbf{s},x_A)T_{B}(\mathbf{s}-\mathbf{%
	\hat {b}},x_B)\times  
\end{equation}
\[\ \times \delta\left( x_A+x_B- e^{-y} \right)d^{2}s  \]
If one uses the standard formula for the number of collisions
\begin{equation}
\left\langle N_{coll}\left( \mathbf{\hat{b}}\right) \right\rangle =ABT_{AB}(\mathbf{\hat{b}}%
)\sigma _{inel}^{qq}\text{,}  \label{eq.8}
\end{equation}
with  Bjorken scaling and a conformal cross-section
\begin{equation}
  \sigma_{inel}^{qq} = \frac{1}{p_T^2} \eqcomma x_{A,B} = \frac{p_T e^{\pm y}}{\sqrt{s}}
    \end{equation}
parallels, the $k_T$ factorization formula (derived through somewhat different physics \cite{avsar}) but reminiscent of a simplification of Eq. \ref{multgen}
\[\
\frac{dN}{ dy} \sim \int d^2p_T  \frac{1}{p_T^2} \int  k dk  \left[ A(x_A)  \tilde{T}_A(x_A,k) \int dx_{A,B} B(x_B) \right. \]
\begin{equation}
\label{kt}
\left. \left.  \tilde{T}_B(x_B,p_T-k) \delta\left( x_A +x_B-e^{-y} \right)\right]\right|_{x_{A,B}=\frac{p_T e^{\pm y}}{\sqrt{s}}} ,
\end{equation}
and $A(x_A),B(x_B)$ are the absolute numbers of partons sitting in that $x$ bin.

Now, according to assumption (a) $\tilde{T}$ is a Fourier-transformable function (the Fourier transform of the nuclear transverse size, which serves as a proxy for the scattering cross-section) characterized by a size parameter $Q_s$
\begin{equation}  \tilde{T}(x,k) \sim \tilde{F}\left( x,\frac{k}{Q_s(x)} \right) \eqcomma Q_s(x) \sim x^{-\lambda}
\end{equation}
Performing all momentum integrals, and using the unitarity assumption (b) we will get, up to a constant
\[\  \frac{dN}{dy} \sim x_A f(x_A) \simeq e^{-y} f(e^{-y})  \]
where the last approximate, due to (c), gives rise to limiting fragmentation.

This is more or less how \cite{jamal,stasto} derived limiting fragmentation in the Color Glass.  However, the $\simeq$ signs should be examined in more detail when different system sizes as well as energies are compared, since generally in the CGC scenario $f(x,Q_s(x,N_{part}))$ will maintain a residual $N_{part}$ dependence, breaking the scaling of $\observable$.

\begin{figure*}[t]
      \epsfig{width=0.99\textwidth,figure=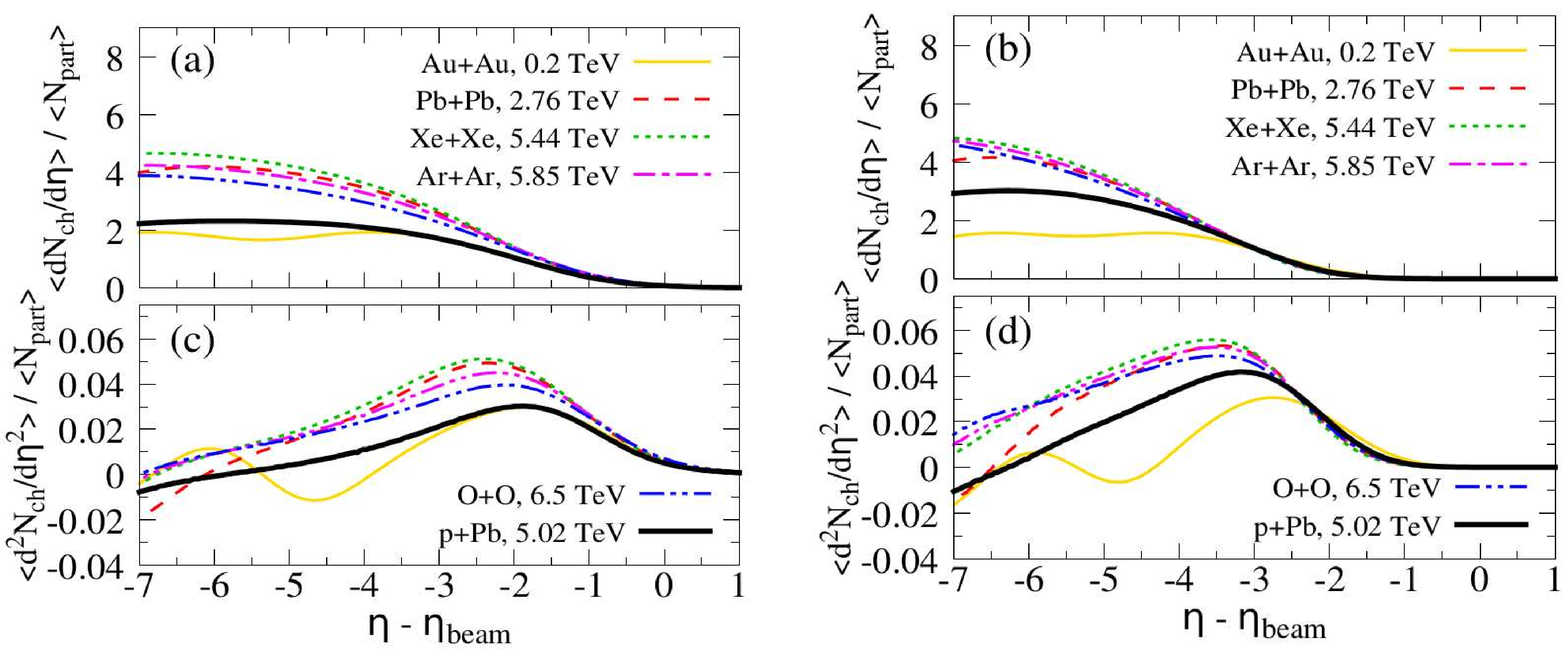}
\caption{\label{cgc} (color online)
$dN/d\eta$ and  $\observable$ calculated in KLN (panel (a),(c)) and rcBK (panel (b),(d)) scenarios.  The normalization was adjusted to reproduce the data in \cite{alice276} using the usual parton-hadron duality parameters}
\end{figure*}

To quantify this effect within the saturation scenario, we have plotted this variable for the rcBK model developed in \cite{andre2}, where the non-linear gluon evolution is solved numerically, as well as the KLN model \cite{kln} which paramterizes the qualitative features of this evolution extrapolating from mid-rapidity.
The normalization was performed with the data of the centrality dependence of charged particle multiplicity at mid-rapidity at 2.76 TeV \cite{alice276}.
The result is plotted in Fig. \ref{cgc}.   As can be seen, the dynamics is qualitatively similar to scenario (b) of Fig \ref{limfraglhc}, with the normalized rapidity density smoothly reaching the universal limit.  In rcBK this limit is reached slightly sooner than in KLN, although the scaling is not perfect in either.  This confirms earlier results published on this topic \cite{jamal,stasto}, as well as predictions in \cite{jeon} (Fig. 8).

However, one should keep in mind that the reason is that the rcBK results reach to the $\eta_{beam}$ region is conditional on the numerical limitation in the rcBK case, since the gluon distribution function has to be extrapolated for $x > 0.01$, which is exactly the kinematical regime where one of the hadrons is probed in the fragmentation region, where a significant quark admixture is present.
It is reasonable to believe, however, that an additional admixture of sources not present at mid-rapidity would weaken the scaling.
In addition, of course, the $k_T$ factorization scenario of Eq. \ref{kt} and saturation in general are, by an order of magnitude estimate, likely to break down at lower RHIC energies, where experimentally limiting fragmentation works quite well (as Fig. \ref{limfragexp} and the accompanying discussion showed), and, for that matter, the calculation of Fig. \ref{cgc}, if taken ``literally'', seems to reproduce this fact.

\begin{figure}[t]
\begin{center}
      \epsfig{width=0.3\textwidth,angle=-90,figure=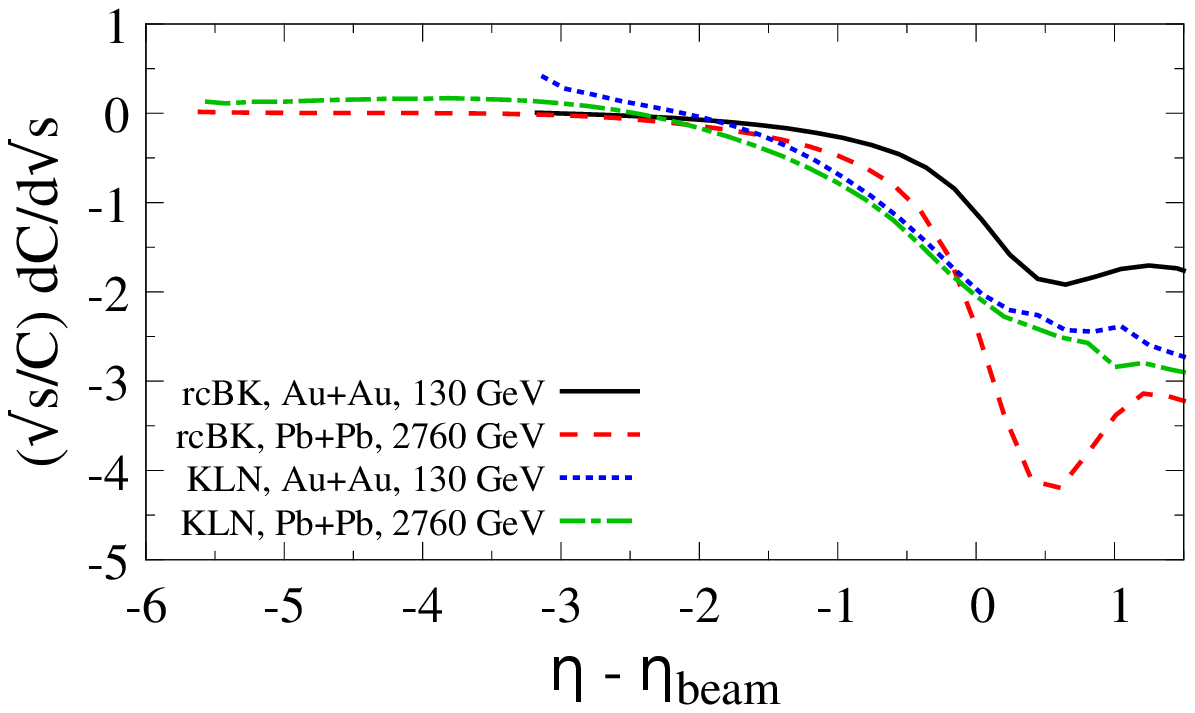}
\caption{\label{cgcC} (color online)
  Violation of limiting fragmentation quantified in the way of Fig \ref{fdelta} and \ref{limfragexp} in the KLN and rcBK models}
\end{center}
\end{figure}

In this spirit, the rcBK calculation done here should  be considered as an ``lower limit'' to limiting fragmentation violation within the CGC scenario, in the
sense that this calculation is likely to be rather inaccurate at many energies and rapidities, but deviations from it are likely to break limiting fragmentation.   And, comparing different system sizes LHC and bottom RHIC energies of Fig. \ref{cgc}, even in the ideal scenario the breaking of limiting fragmentation would be substantial.
Putting the results of Fig. \ref{cgc} into the form of Fig. \ref{fdelta} and \ref{limfragexp} is shown in Fig. \ref{cgcC}.   We confirm that the deviation from limiting fragmentation in these models is higher than the data, and surprisingly also of the Glauber model when normalized.
So, what conclusions would we draw if it is found to hold, quantitatively, for pA, dA and AA collisions up to LHC energies?   Given the success of the wounded quark model, we could combine it with partonic limiting fragmentation insights to
  calculate rapidity-dependent ``wounded partons'', the number of partons that experienced at least one collision.  Without resort to non-perturbative physics, it is difficult to justify this model, since particle production is thought to be perturbative in the regime where partons are good degrees of freedom, and perturbative dynamics does not scale with participants.
That said, since wide rapidity intervals correspond to wide spacetime rapidity intervals the idea that confinement effects seep into the longitudinal evolution of partons, and hence ``wounded partons'' become relevant, is not so unreasonable.
Since, phenomenologically, a wounded quark model appears to work \cite{woundq1,woundq2,woundq3,woundq4} well enough that fusing this with the geometrical deep inelastic scattering warrants a try.

Using the usual formula for the number of wounded nucleons adapted to the wounded parton picture, with the Fourier transform of the transverse participant density of nucleus $A$
\[\
 \tilde{N}_{part}^{qA}(x_A,k_A,A,B) \simeq  A(x_A) \tilde{T}_{A}(k_A,x_A) \int  d^2 k_B dx_B  \times
\]
\begin{equation} \times  \left( 1- \left(1-\frac{\tilde{T}_{B}\left(k_A-k_B,x_B\right)}{\left(k_A-k_B\right)^2}   \right)^{B(x_B)} \right) \end{equation}
If one introduces  the emission function \cite{woundq5} for a wounded parton characterized by $x_A,k_A$ to emit a particle of $p_T,y$ as $F(p_T,y,k_A,x_A)$ one can get the multiplicity in terms of the wounded partons.

 For each nucleon collision we will have
\begin{equation}\label{eq16}
\frac{dN}{dy} = \int p_T dp_T dk d x F(p_T,y,k,x) \times
\end{equation}
\[\  \times \left( \tilde{N}_{part}^{qA}(x,k,A,B) + \tilde{N}_{part}^{qB}(x,k,B,A) \right) \]
Expanding to first order in $\tilde{T}_{B}(k)/k^2 $ it is possible to reduce this expression into a sum of terms of the form
\[\ \frac{dN}{dy} \sim \int p_T dp_T \int dx_{A}  k_A dk_A   A(x_A) \tilde{T}(k_A,x_A) \times  \]
\[ \times  F(p_T,y,k_A,x_A)   \int dx_B k_B dk_B \frac{B(x_B)\tilde{T}_{B}(k_A-k_B,x_B)}{\left(k_A-k_B \right)^2}  \]
assumption (b), together with the reasonable requirement that $B(x\rightarrow 0)$ does not diverge will remove the second integral to an approximate constant value for each nucleon.
Assumption (c) ensures $F(p_T,y,k,x) \simeq F(p_T,y,k,e^{y})$ and also that $k_A-k_B \simeq k_A$.  Additionally assuming the equivalent of (b) and (c) for  $F(...)$, i.e. that it is only a narrowly peaked and normalized function of $y$ differences,  will ensure the resulting function will only depend on $e^{-y}$ analogously with Eq. \ref{kt}.  Hence, limiting fragmentation is recovered.

Unlike the Color Glass, however, one can straight-forwardly extend limiting fragmentation to $AA$.  What prevented limiting fragmentation in the wounded nucleon model was the fact that $f=1$ in Eq. \ref{fpar} is incompatible with any choice of $\alpha$ in equation \ref{alphadef} that fits multiplicity at mid-rapidity.
Physically wounded partons, by allowing participants to be spread across $x$ and nuclear size to vary across $x$, make it possible to combine a purely participant $f=1$ scaling (which does not depend on $\sqrt{s}$) with strongly non-logarithmic dependence of multiplicity per participant, obtained through variation with $x$ of $A(x)$ and $\tilde{T}(x)$.   The difference with the Color Glass is that $\tilde{T}(x)$ would still be individual nuclear collisions, albeit with a size that effectively depends in Bjorken $x$.   Thus, one expects $Q_s(x)$ (``s'' is size and not saturation here!) to be independent of the number of participants, leading to a universal $\observable$.

  Previous literature on wounded quarks \cite{woundq1,woundq2,woundq3,woundq4,woundq5} considers nucleon transverse substructure but does not separate it into $x-$distributions.  This allows the authors to fit mid-rapidity data or rapidity data at a given energy.  Generalizing these models along the lines presented in this work while maintaining the current phenomenological agreements seems like a straight-forward exercise, albeit one beyond the scope of this paper.
Thus, if an eventual confirmation of a  $\observable$ universal across energies and system sizes will indeed be observed, the wounded parton model will be a promising avenue to model this.
\section{discussion and summary}
In this work, we have examined the behavior of $\observable$, the multiplicity density per participant, at high rapidity, motivated by the near independence with energy of its first derivative.   The universal behavior of $\observable$ was found to generally break in a Glauber model, particularly in the two-component models usually used to fit data at mid-rapidity.   Thus, this is a good observable to constrain such models.

The reason these models fail is physically very simple:  Inasmuch as both collisions and participants contribute to multiplicity, the energy dependence of the two factors is non-trivial and generically different from the rapidity dependence.   This means that the height of the distribution depends non-trivially on the energy while the width is constrained by kinematics.   Absent unnatural cancellations, limiting fragmentation should be broken.

The only way to restore it is to make multiplicity dependence entirely driven by wounded nucleons at all energies.  This is however not enough, since the price for this is to make the multiplicity per nucleon in $AA$ collisions rise with energy much faster than logarithmically.  This, as well as clashing with experimental data below LHC, will generally break limiting fragmentation as well.
A ``wounded parton model'' might be able to evade such a constraint since there the ``size of the wounded degree of freedom'', instead of being encoded in the cross-section, is allowed to vary with longitudinal $x$, which in this case is tightly correlated with momentum rapidity.   Energy and rapidity distributions are therefore naturally correlated in this regime.     The observation of a universal $\observable$ could be indicative of ``wounded quark'' dynamics.

One can ask how universal is the class of models reproducing such universality.
Models such as the Color Glass \cite{stasto} share some similarities with the wounded parton scenario considered in the previous section, but there nuclei lose their individuality and $Q_s$  is common to the same area of transverse space (``s'' in color glass models is saturation instead of size!).   Also, Eq. \ref{kt}($k_T$ factorization) used to connect the gluon density to particle density in this regime leads to a universal slope in the fragmentation region in the same way as Bjorken scaling.

Indeed, as \cite{stasto} finds approximate limiting fragmentation,
explained by the  factorization of parton distributions in target and projectile at large rapidities together with the fact that the multiplicity distribution is directly proportional to the parton
density in the target and relatively independent of the scales of the process.
The wounded parton model conjectured in the previous section shares these characteristics.
We note, however, that \cite{stasto} predicts some violation of limiting fragmentation inasmuch as the assumptions above cease to have validity.
Thus, the calculations of \cite{stasto} point to some violation of $\observable$, which turns out to be comparable, if not larger, than the Glauber model.

Looking at \cite{schee1,schee2,groz} a similar discussion can be made about AdS/CFT initial states, where the breaking of scaling appears even stronger as it is controlled both by a critical transparency and the coupling constant.
Thus, some breaking of limiting fragmentation, when all energies and systems sizes are concerned, appears likely in all models claiming connection to field theory (the wounded parton model so far does not).

We continue with experimental considerations.   To our knowledge, so far
measurements at high enough rapidity to compare to even top RHIC energy were not done, with the closest experimental measurement being \cite{alicey,cmsy,atlasy}.  Also, a result, seeming to confirm scenario (b) of Fig. \ref{limfraglhc}, was obtained for $dE_T/d\eta$ by the CMS collaboration using the CASTOR detector \cite{castor}.
Since multiplicity and transverse energy have a non-trivial separate dependence which is also sensitive to system size \cite{alicept}, we hesitate at drawing conclusions there.
 \cite{alicephoton}, in contrast to CASTOR, has reported a breaking of limiting fragmentation in inclusive photons.   While QED processes are not expected to limiting fragment (there is no hadronization, and $y \sim \ln (1/x)$ is not expected to hold), some 85$\%$ of photons in \cite{alicephoton} are thought to come from $\pi^0$ decays \cite{cosentino}.   Hence, this result makes a breaking of limiting fragmentation at LHC energies likely.   In summary, until a direct measurement of both $dN/d\eta$ and $dE_T/d\eta$ is performed, relying on these data to make a conclusive statement is difficoult.

Studies comparing the rapidity dependence of p-Pb and Pb-Pb are totally lacking.
Comparing with smaller asymmetric systems, such as Pb-Pb and p-Pb collisions, where a deviation from purely wounded dynamics should be more pronounced the observable can be studied on the ``same side'', as discussed in \cite{jeon}.

The fact that most experiments focus the detector on mid-rapidity of course makes this measurement problematic.  We want to point out, however, that this is a bulk observable, not requiring particle identification or momentum measurement, problematic since particles at high rapidity are highly relativistic.  The existing small experiments at rapidity comparable to lower RHIC energies\cite{totem,lhcf} as well as LHCb \cite{lhcb} could take part in this investigation together with the larger collaborations.

Since the calculations here were focused on proof-of-concept estimates testing for violation, we will give a ``cartoon'' of what we expect in each scenario, with the  the alternatives are summarized in Fig. \ref{limfraglhc}.  The key is to go to a high enough rapidity as to compare with a lower energy.   If limiting fragmentation still holds, $\observable$ will evolve to smoothly ``touch'' the corresponding value at mid-rapidity of that energy (scenario (b) of Fig. \ref{limfraglhc}).  Otherwise, the slopes will be different.

In this paper, we have limited ourselves to multiplicity, which, given a low-viscosity nearly isentropic fluid evolution, can be considered to be an initial state effect \cite{molnar}.  However, past experimental results also reported to have seen limiting fragmentation for elliptic flow at RHIC energies \cite{phobosrev}.  This is considered to be a final state effect, sensitive primarily to the transport coefficients and freeze-out dynamics of the system \cite{weakly}.   Should limiting fragmentation of flow observables, or even of average transverse momentum, be confirmed at the LHC\footnote{ We note in passing that the rapidity dependence of elliptic flow in pA,dA and AA collisions measured in \cite{dav2}
is qualitatively what you would expect given a universal $v_2(p_T)$ curve together with a dependence of average momentum on multiplicity and system size expected from \cite{alicept}}, especially in events of same eccentricity but different size, one might have to rethink this paradigm and start exploring scenarios where ``flow'' arises as an initial state effect \cite{gambini,raju,bierlich} in the systems concerned.

In conclusion, we have examined limiting fragmentation in various phenomenological models, namely Glauber, Color Glass, and wounded quarks.   Glauber models generally fail to reproduce limiting fragmentation at LHC energy once they were tuned to reproduce LHC data.   The same is true for Color Glass models once different system sizes are considered.  We have further argued that wounded parton scenarios have the potential to model limiting fragmentation also at LHC energies and for all system sizes.   Since such limiting fragmentation has not to date been verified, no such model can be considered to have been ruled out.
Rather, this paper motivates an experimental search for this observable, and generally for a comparison between low-energy bulk observables and the high rapidity limit of the same observable at high energy.   We eagerly away these experimental results.

\textit{Acknowledgements}   GT acknowledges support from FAPESP proc. 2017/06508-7,
partecipation in FAPESP tematico 2017/05685-2 and CNPQ bolsa de
 produtividade 301996/2014-8. DM was supported by CNPQ graduate fellowship n. 147435/2014-5. This work is a part of the project INCT-FNA Proc. No. 464898/2014-5.  We thank Fernando Navarra, Radoslaw Ryblewski, Mauro Rogerio Cosentino and Wit Busza for fruitful discussions and suggestions.
A.V.G. acknowledges the Brazilian funding agency FAPESP
for financial support through grants 2017/14974-8 and 2018/23677-0.

\appendix
\section{A note on rapidity and pseudorapidity}
        \begin{figure*}[t]
\begin{center}
      \epsfig{width=1.\textwidth,figure=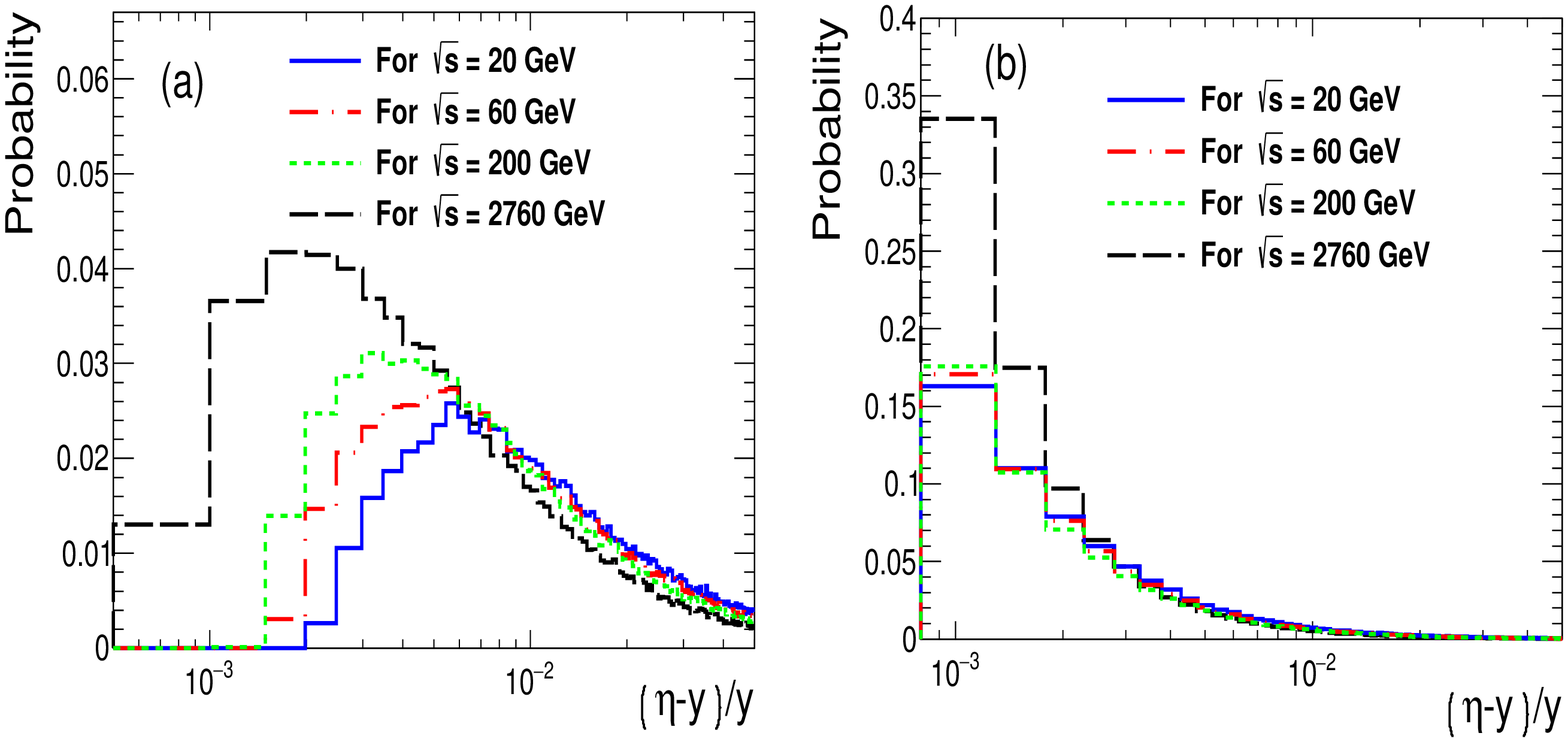}
\caption{\label{yeta} (color online)
  Estimated distribution of $(y-\eta)/y$ for protons (panel (a)) and $\pi$ (panel (b)) assuming $\ave{p_T}$ does not deviate from its mid-rapidity value.}
  \end{center}
    \end{figure*}
  We note that the experimental observable usually measured here is not rapidity (the $z$ direction, where momentum is $p_z$.   The total momentum is $p$)
  \begin{equation}
 y = tanh^{-1} \left( \frac{p_z}{\sqrt{p^2+m^2}}  \right)
    \end{equation}
  but the pseudo-rapidity, a function of the angle with the beam axis which does not require particle identification
      \begin{equation}
 \eta =   tanh^{-1} \left( \frac{p_z}{p}  \right)
    \end{equation}
    The rapidity is the observable with the ``nice`` transformation property of being linear under boosts.  It is also related to Bjorken $x$ by eq. \ref{ygdef} and equal to the space-time rapidity
 in the Boost-invariant limit (Eq. \ref{spacetime}).  However, it is very difficoult to measure for ultra-relativistic particles as it requires particle identification, and/or the simultaneus measurement of the energy and momentum.

    The pseudo-rapidity has no ``nice`` properties, but it is very easy to measure as it is directly related to the longitudinal angle $\theta$
    \begin{equation}
\eta =-\ln \tan \left(\frac{\theta}{2} \right)
      \end{equation}
    At mid-rapidity, binning the distribution in terms of $\eta$ rather than $y$ causes the near-gaussian distribution measured in \cite{whitebrahms} to aquire a ''plateau'' necessitating a double gaussian fit, as was done in \cite{ampt}.

    Away for mid-rapidity ,
    for the great majority of produced particles, the two are interchangeable,
    \begin{equation}
\eta -y \sim - \order{ \frac{m^6 p_z^3}{p^9}} \simeq \frac{\cos\theta}{2}\left( \frac{m}{p_T}\right)^2
    \end{equation}
    In this work we do not fit the rapidity distribution globally, but rather rather concentrate on observables local in rapidity sensitive to limiting fragmentation: 
    the slope in the fragmentation region
        \begin{equation}
          \abs{y} \geq \order{\frac{1}{2}-\frac{3}{4}} \abs{y_0}
        \end{equation}
        and its derivative w.r.t. $\sqrt{s}$.
Hence, we assume differences between $y$ and $\eta$ to be negligible.
    
    To estimate the goodness of this assumption, lacking particle identification and $p_T$ measurements away from mid-rapidity, we shall
    take a limiting fragmentation inspired rapidity distribution \cite{jeon}
    \begin{equation}
 \frac{dN}{p_T dp_T dy} =C \left( \ln \left[ \frac{ \sqrts}{m_p} \right]-y \right)  \exp \left[- \frac{p_T}{2 \ave{p_T}} \right]
    \end{equation}
    where
\[\    \ave{p_T}= \left\{  \begin{array}{cc}
        500 \mathrm{MeV} & \pi\\
        1 \mathrm{GeV} & p
        \end{array}
      \right.
    \eqcomma C \simeq 0.65  \]
    and histogram $(\eta-y)/y$.   The results are shown in Fig. \ref{yeta} 
for protons and pions, in the region of rapidity considered in the rest of the paper, $y_0  -y \leq 6$ or so (the figure should be symmetrized if both ``target`` and ``projectile`` are considered) and assuming a thermal transverse $p_T$ distribution with $\ave{p_T}$ weakly dependent on rapidity (central values are as in \cite{starscan}).    
As can be seen any reasonable admixture between baryons and mesons will result in a systematic error of order of a percent if $\eta$ and $y$.  This error is significantly lower than other experimental errors, and hence is neglected in this work.

    To correct this would require a rapidity as well as the energy dependence of the temperature, baryo-chemical potential and $\ave{p_T}$, something currently unknown 
(measured very partially in \cite{whitebrahms} and \cite{starscan}), but, in the spirit of the rest of the paper, any non-trivial variation of this is likely to spoil limiting fragmentation.

The observables treated in this paper, defined around the second derivative of the rapidity distribution away from mid-rapidity, are optimized to minimize the effect of the rapidity-pseudo rapidity interchange.   Alternatively, for example, analyzing the width of the rapidity distribution as a whole (as was done, for example, in \cite{ampt}) is much more sensitive to this distinction, since the mid-rapidity plateau varies significantly between $\eta$ and $y$, and this influences both the height the width of the distribution in a particle-dependent manner (see \cite{becrap} for a discussion of this issue at low energies).

\end{document}